\documentstyle[12pt]{article}
\makeatletter

\ifcase \@ptsize
\or 
\or 
\fi
\advance\textheight by \topskip

\ifcase \@ptsize
\or 
\or 
\fi

\def\section{\@startsection {section}{1}{\z@}{-3.5ex plus -1ex minus 
 -.2ex}{2.3ex plus .2ex}{\large\bf\centering}}
\def\subsection{\@startsection{subsection}{2}{\z@}{-3.25ex plus -1ex minus 
 -.2ex}{1.5ex plus .2ex}{\sc}}
\def\@cite#1#2{\nolinebreak$^{[\scriptstyle #1\if@tempswa , #2\fi]}$}
\def\@citex[#1]#2{\if@filesw\immediate\write\@auxout{\string\citation{#2}}\fi
  \def\@citea{}\@cite{\@for\@citeb:=#2\do
    {\@citea\def\@citea{,\penalty\@m}\@ifundefined
       {b@\@citeb}{{\bf ?}\@warning
       {Citation `\@citeb' on page \thepage \space undefined}}%
{\csname b@\@citeb\endcsname}}}{#1}}
\@definecounter{footnote}

\gdef\@publabel{\hfil}
\gdef\@pubdate{\null}
\gdef\@pubnumber{\null}
\gdef\@author{\null}
\gdef\@title{\null}
\gdef\@abstract{\null}
\long\def\pubdate#1{\gdef\@pubdate{#1}}
\long\def\pubnumber#1{\gdef\@pubnumber{#1}}
\long\def\publabel#1{\gdef\@publabel{#1}}
\long\def\author#1{\gdef\@author{#1}}
\long\def\title#1{\gdef\@title{#1}}
\long\def\abstract#1{\gdef\@abstract{#1}}
\def\titlerelax{
\let\maketitle\relax
\let\settitleparameters\relax
\let\consolidatetitle\relax
\let\inittitlepage\relax
\let\finishtitlepage\relax
\let\titlepagecontents\relax
\let\multithanks\relax
\let\titlebaselines\relax
\let\@makepub\relax
\let\@maketitle\relax
\let\@makeauthor\relax
\let\@makeabstract\relax
\let\@maketitlenote\relax
\let\thanks\relax
\let\titlerelax\relax}
\def\titleclean
{\gdef\@titlenote{}
\gdef\@abstract{}
\gdef\@author{}
\gdef\@title{}
\gdef\@pubdate{}\gdef\@pubnumber{}\gdef\@publabel{}
\gdef\@dpublabel{}
}

\def\@makepub{\vbox to \z@{\hbox to \textwidth{\hfill
\@publabel \hfill
\llap{\parbox[t]{0.25\textwidth}{\raggedleft\@pubnumber}}}%
\vss}}
\def\@maketitle{\vskip 60pt \begin{center}
 {\LARGE \@title \par}
 \end{center}}
\def\@makeauthor{{\def\and{\smallskip {\normalsize \rm and\smallskip}}
\long\def\address##1{{\def\and{\\and\\}\medskip
				{\small \it \\##1\\}
}}
{\centering
 \vskip 1.5em
 \large \lineskip .75em
 \@author}
 \par}} 
\def\@makedate{\vskip 1.5em 
 {\raggedright \small \noindent\@pubdate \par}}
\def\@makeabstract{\vskip 1.5em
{\small 
\begin{center}
{\bf ABSTRACT\vspace{-.5em}\vspace{0pt}} 
\end{center}
\quotation \@abstract \endquotation}
\vspace{-1em}}
\def\maketitle{
\let\footnotesize\small \setcounter{page}{1}
\@makepub
\@maketitle
\@makeauthor
\@makeabstract
\@thanks
\@makedate
}

\begin{document}
\newcommand{\hsp}{\hspace{0.08in}}
\newcommand{\eqn}{\begin{equation}}
\newcommand{\ee}{\end{equation}}
\newcommand{\be}{\begin{equation}}
\newcommand{\bt}{\beta}
\newcommand{\al}{\alpha}
\newcommand{\eeaa}{\end{eqnarray*}}
\newcommand{\beaa}{\begin{eqnarray*}}
\bibliographystyle{npb}

\pubnumber{DAMTP-97-129 \\ 
hep-th/9711148 \\
November 97}

\title{Quantum affine Toda solitons}
\author{N. J. MacKay\footnote{\tt n.j.mackay@damtp.cam.ac.uk} 
\\[0.1in]
{\small\em contributed talk at the 12th International Congress 
of Mathematical Physics,\\ Brisbane, July 1997}
\address{Dept of Applied Maths and Theoretical Physics, \linebreak
         Cambridge University, \linebreak
         Cambridge, CB3 9EW, UK}}
\abstract{We review some of the progress in affine Toda field theories
in recent years, explain why known dualities cannot easily be
extended, and make some suggestions for what should be sought instead.}

\maketitle
\baselineskip 18pt
\parskip 12pt
\parindent 10pt

\section{A brief review}

The affine Toda field theories are models of real scalar fields,
in one space dimension (here taken to be the real line\footnote{The 
theory on the half-line is another story\cite{corr96}.}), with
exponential interactions. The Lagrangian is 
\be\label{L}
{\cal L}={1\over2}(\partial_\mu \phi)(\partial^\mu\phi)-{m^2\over\bt
^2}\sum_{j=0}^n n_j(e^{\bt\al_j\cdot\phi}-1)\;, 
\ee
where the field $\phi(x,t)$ is an $n$-dimensional vector, $n$ being the rank
of the finite Lie algebra $g$. The $\alpha_j$ ($j=1,...,n$) are
the simple roots of $g$; $\alpha_0$ is the lowest root, so that
\{$\al_0,\al_j$\} are described by one of the extended Dynkin
diagrams of an affine algebra $\hat{g}$. 
The fields can be rescaled so that $\bt$ only appears in ${\cal L}$ 
through an overall factor $1/\bt^2$; expanding in powers of $\bt^2$ 
is thus equivalent in the quantum theory to expanding in $\hbar$.

A great deal has been discovered about these models in the last ten
years, and the resulting spectra of perturbative and solitonic
particles are surprisingly rich. It is clear, however, that vastly more
remains to be learned about the setting for these results. This paper
does not provide a pedagogic review: excellent introductions 
already exist\cite{corr94,gandethesis}, and the reader
is referred to these for full details of
masses, S-matrices and so on. Rather I shall provide a brief
qualitative summary, and then describe the current frontier.
In particular I want to explain why there appears to be no duality
among the full spectra of solitons and particles.

\subsection{Real $\bt$, simply-laced $g$}

The Lagrangian (\ref{L}), with $\bt$ real, is amenable to
simple techniques. The vacuum $\phi=0$ is unique; the 
exponentials may be expanded about it, and
the masses (of order $m$) and the three-point couplings of the mass
eigenstates may then be read off\cite{brade90,chris90}. 
The masses form an eigenvector of
the Lie algebra's Cartan matrix\cite{free91}, 
and so may be put into correspondence
with spots on its Dynkin diagram - we shall refer to this as the
particle's `species'.

The theory is further known to be classically integrable,
with local, commuting conserved charges of spins equal (for $g$
simply-laced) to the exponents of the Lie algebra\cite{wilson80}: 
the first
exponent is always one, corresponding to energy-momentum.
Their conservation constrains the admissible
three-point couplings very strongly, indeed to such an extent
that it might be surprising that solutions exist at all.
Nevertheless an admissible set does exist for each algebra, 
as is made beautifully clear in the construction due to 
Dorey\cite{dorey91,fring91}. 
When two particles fuse to form a third, the conservation of each 
charge requires the existence of a triangle with sides of lengths 
equal to the charges and angles determined by the particles' momenta. 
In Dorey's construction, each three-point coupling corresponds to a 
triangle of roots, each being chosen from an orbit of a root under a
Coxeter element of the Weyl group. The projection of this equation in 
the (higher-dimensional) root space onto various planes, on which the 
Coxeter element acts as a rotation, gives the triangles for the
charges.

These results at tree-level can be quantized perturbatively; and
at one-loop order, for simply-laced $g$, the particle masses all
renormalize in the same ratio. The expectation is that this will
be true to all orders, and exact S-matrices (which, since the
particles are scalar, will be scalar factors, in fact products
of simple trigonometric functions) may be hypothesized for 
the particles\cite{arin79,brade90,chris90},
with rigid mass ratios and hence pole structure, described by Dorey's
rule. Further, the dependence on $\bt$ is such that there
is a strong$\leftrightarrow$weak coupling duality: the S-matrices
are invariant under $\bt\mapsto 4\pi/\bt$.

\subsection{Imaginary $\bt$, simply-laced $g$}

If in the simplest case, $g=a_1$,  we take $\bt$ to be imaginary,
we have moved from the sinh- to the sine-Gordon model. The
vacuum is now degenerate, and there exist solitons, with 
masses proportional to $ m/\bt^2$, and `breathers,' consisting of an
oscillating soliton-antisoliton pair (at imaginary rapidity
difference), with a continuous mass spectrum. Exact S-matrices, now 
with matrix structure, may be proposed for the solitons,and the bound 
states include a discretized spectrum of breathers. The perturbative 
particle is still present, and its mass and S-matrix are equal to 
those of the lowest breather state\cite{zam79}. It is therefore
natural to suspect that in the exact quantum theory the breather 
and the particle are identical.

For other $g$, however, ${\cal L}$ is complex for imaginary $\bt$.
Nevertheless there exist solitons\footnote{Issues of reality 
and stability are still in doubt\cite{hollo93b,evans92,khast95}.}, 
constructed by finding Hirota $\tau$-functions\cite{hollo92} or 
with vertex operators\cite{olive93}, of each species and with
topological charges in (but not filling) the corresponding fundamental
representation\cite{mcg94}, interpolating the (now degenerate)
vacua. They have real masses, of order $1/\bt^2$ and proportional
to those of the particles, and real higher conserved
charges\cite{free94}, and they `fuse' - double solitons truncate
and appear as single solitons - according to Dorey's rule. 
They also obey an exclusion rule: static multisolitons exist,
but only if no two of the constituent single solitons are of the
same species. There are also breathers in every species, and excited 
or `breathing' solitons in some\cite{harde94}; these have the 
expected continuous mass spectra.

Any quantization of such an apparently non-unitary theory
must be very suspect, but it is possible to plough ahead with,
for example, semiclassical quantization, and obtain sensible results:
that, as with the particles, the soliton mass ratios do
not renormalize. The expectation is that, in a sense yet to be
discovered, there is an embedded unitary theory, of solitons and their
bound states, within the larger non-unitary one\cite{spence95}. 
Whether such a unitary theory can be defined for all values
of the coupling constant is more equivocal. Finally it is
worth pointing out that, in contrast to the
sine-Gordon$\leftrightarrow$massive Thirring model relationship,
there is no known quantum field theory of which the affine Toda
solitons are the elementary excitations, though the particle
S-matrices have been expressed as exchange relations\cite{corr91}.

As with sine-Gordon solitons, we can try to
construct exact soliton S-matrices. Traditionally the route
used has been the implication from conservation of the local charges
that multiparticle scattering must factorize into products of 
two-particle scattering processes, as described by the Yang-Baxter 
equation (YBE). However, it is now recognized that underlying the YBE is
a set of conserved charges forming a quantum group\cite{berna90}. 
In contrast to the charges described earlier, these are non-local: 
that is, if measured on two 
asymptotically separate physical states, instead of yielding
simply the sum of the values of the charges on the individual
states, there are further interaction terms.  Further, they 
do not commute - in fact they form a quantum group - and they
have indefinite or non-integer spin. 
The multiplets in the theory will form representations of them
(scalar for the particles, non-scalar for the solitons), and their
conservation can be used to determine the S-matrices up to a scalar
factor.

In the affine Toda theories the charges have non-integral spins of
order $4\pi/\beta^2-1$, and form a quantized affine
algebra \cite{feld91} in the principal gradation, with deformation 
parameter $q=e^{4i\pi^2/\beta^2}$.  There is a
(topological charge-dependent) similarity transformation to the 
homogeneous gradation, in which the spectral parameter $x$ is 
$e^{(4\pi/ \beta^2-1)\theta}$
(where $\theta$ is the incoming particles' rapidity difference).
It is important to distinguish this realization of the quantum group
from that obtained by quantizing the auxiliary algebra of the Lax
pair, where the deformation parameter is $q=e^{i\beta^2\over 2}$,
and which thus has the conventional classical ($\beta\rightarrow 0$)
limit, the undeformed algebra. 
In contrast, in the classical limit of the S-matrices, whilst the 
quantum charge algebra itself becomes highly deformed, the ambiguities 
(multiples of $2i\pi$ in the exponents) in $\theta$ at the poles 
(which occur at $x$ equal to
some power of $q$) become relatively small, and the discrete bound
state spectrum becomes continuous, as expected. 

The two forms of $q$ are very suggestive of some strong-weak coupling
relationship between the auxiliary and quantum spaces. This is not a
phenomenon only of affine Toda theories: it also occurs with
Yangians\cite{macka95}, where there is a similar $\hbar\leftrightarrow
1/\hbar$ relationship. It is unclear how or whether to attach any 
significance to this, relating as it does a mathematical artefact
(the auxiliary algebra) to physical charges (the quantum algebra).

The physical states are expected to fall into fundamental representations 
of the quantized affine algebra. These
are reducible representations of the ($q$-deformed) Lie subalgebra,
containing its corresponding fundamental representation as a
component. If we believe (as all the classical and semiclassical
information leads us to) that these multiplets are quantum solitons,
an oustanding issue is how they succeed
in filling representations which are classically only part-filled.
Pressing on, we encounter the outstanding open problem of the
general construction of the relevant solutions
of the YBE. The only simply-laced case for which it is solved
completely is $a_n$, and the results are precisely what one might 
hope for\cite{hollo93,gande95,gandethesis}. If we apply the bootstrap
principle, that all poles are to be interpreted in terms of on-shell
diagrams of physical particles, and that, when the diagram is 
tree-level, S-matrices for the intermediate state can be constructed 
by fusion, the spectrum needed to close the bootstrap consists of 
the solitons and discretized spectra of breathers
and breathing solitons. The analysis of the particles for real $\beta$
can still be carried out for imaginary $\beta$, and their masses and 
S-matrices are the same as for the lowest breathers, leading us to
identify them.

Even in the absence of explicit S-matrices for other $g$, 
something
can still be said about their expected pole structure. Mathematically, 
the S-matrix fusings are the analogue for the quantum group of the
Clebsch-Gordan rule for Lie algebras (to which Dorey's rule
is related\cite{brade91}, but not identical): they are its rule
for tensoring representations. For the fundamental representations,
this rule has recently been shown to be precisely Dorey's 
rule\cite{chari95}.  The proof is exhaustive, however,
so its truth is very suggestive of unseen mathematical structure
beyond our current understanding of quantum groups. Mirroring
these facts about the algebraic structure is a general observation
that the scalar prefactors of the expected S-matrices are related
to the `quantum dilogarithm'\cite{johns97}. 
A relationship between two such different pieces of mathematics would
come as no surprise: as we have seen, ATFTs have many ways of
exhibiting Dorey's rule.

\subsection{$g$ nonsimply-laced}

When $g$ is nonsimply-laced, the classical soliton masses are no
longer proportional to those of the particles. In fact, they are
proportional to those of the particles in the dual theory\cite{olive93}, 
based on $g^\vee$, with
the root lengths inverted; we shall call this `Lie duality'. 

In addition to these untwisted algebras,
whose affine extensions we denote $g^{(1)}$ and $g^{\vee(1)}$, 
we must now also consider the twisted algebras $g^{(1)\vee}$, obtained
from a nonsimply-laced $g$ by inverting the lengths of both the 
$\alpha_i$ and $\alpha_0$. (Because of the involvement here of
$\alpha_0$, we shall call this `affine duality'.)
Here, both the particles and the solitons are a subset of those 
for a simply-laced algebra $h^{(1)}$, of which the twisted algebra
is a subalgebra invariant under an automorphism of order $k$:
in the usual notation, $g^{(1)\vee}=h^{(k)}$.

For nonsimply-laced algebras, the particle mass
ratios\cite{brade90,macka94} (computed
to one-loop order) are no longer $\beta$-independent. In fact
the strong$\leftrightarrow$weak duality mentioned earlier extends 
to this case, using affine duality of algebras: the masses\cite{deliu92} and 
S-matrices\cite{deliu92,dorey93,corri93}  of the $(g^{(1)},\bt)$
theory are those of the (twisted) $(g^{(1)\vee},4\pi/\bt)$ theory.

Dorey's rule has not yet been fully generalized to nonsimply-laced 
cases. The rule for $g$ predicts the correct (tree-level)
three-point couplings, but careful examination of the particle
S-matrices\cite{dorey93,corri93}  shows
that not all of these are allowed fusings in the exact quantum theory.
In fact Dorey's rule can be extended to cover twisted 
algebras\cite{dorey94,chari95}, and
it is the intersection (which we shall denote $D(g)$) of the sets 
of couplings allowed by the rules for $g$ and for $g^{(1)\vee}$ which 
gives the correct set. However, a full generalization needs also
to give the correct flexible mass and charge triangles, and, despite
some progress, this has not yet been achieved.

The soliton masses\cite{hollo93b,macka94,deliu94b} (computed 
semiclassically) no longer renormalize in the same ratio either:
in fact the quantum-to-classical soliton mass ratio is, up to
a species-independent factor, that of the particle\footnote{Although
almost certainly true\cite{gande96}, this matter is not settled:
the semiclassical calculations have never been completed entirely
satisfactorily.} Because of
the Lie duality mentioned above, however, there is no
affine duality of soliton masses. This might have been expected:
remember that whereas the particles have masses of order $m$,
solitons have masses of order $m/\beta^2$, heavy in the weak limit,
which would be expected to become very light for strong coupling.
Further, the S-matrices, as can be seen from $x$ and $q$ (whose
form is similar for nonsimply-laced algebras), will look very different
for strong and weak couplings.

In calculating the soliton S-matrices, we encounter the subtlety that 
for the $\hat{g}$ theory, the non-local charges form 
$U_q(\hat{g}^\vee)$\cite{feld91}. Their spins now depend on the
root length, with the algebra in what has been
termed the `spin gradation'. It can still be transformed to the
homogeneous gradation, however, though with a slightly different
form for $x$.

The first complete set of S-matrices
for a nonsimply-laced theory was found\cite{gande95b} 
for the twisted algebra $d_{n+1}^{(2)}$, and contains the 
expected results: solitons, excited solitons and breathers,
with the lowest breather state identifiable with the particle. 
The soliton fusing rule is $D(c_n)$, as expected. 

The most subtle case is that of the untwisted nonsimply-laced
algebras, where
we must construct YBE solutions corresponding to twisted affine
algebras\cite{deliu95b}. This was done\cite{gande95c} for the 
$b_n^{(1)}$ theory, and the classical result mentioned in 
the first paragraph for the particle was found to
apply to the lowest breather state as well: its mass 
is not always proportional to that of the soliton. Thus we
now believe that in all cases the lowest breather is to be identified
with the particle; in support of this, the exact masses and
S-matrices turn out to be identical. 
The soliton fusing rule turns out to be $D(c_n)$, suggesting
a Lie duality (not yet proven in general): that $g$ solitons have a 
$D(g^\vee)$ fusing rule.

\section{Towards an overview}

We can discuss these facts within the following scheme\cite{gande95b}:

\[
\begin{array}{ccc}
g^{\vee(1)},\epsilon \;\;{\rm solitons} &  &
\left(\hsp g^{(1) \vee},{1\over\epsilon'} \;\;{\rm solitons}
\hsp\right) \\[0.2in]
\updownarrow{\scriptsize\rm Lie} & & \downarrow\,? \\[0.2in]
g^{(1)},\epsilon' \;\;{\rm particles} & \hspace{0.1in}
\stackrel{\rm Affine}{\longleftrightarrow} &
g^{(1) \vee},{1\over\epsilon'} \;\;{\rm particles}  \\[0.2in]
\end{array}
\]
Here $\epsilon\equiv\bt^2/4\pi$ where $\bt$ is the real coupling:
thus in the top-left, classical solitons would be obtained when
$\epsilon$ is negative.
Everything in this diagram has the same (exact) mass ratios, 
extracted from the proposed S-matrices, and the same $D(g)$ fusings. 
Solitons are in the top row, particles in the bottom; the left 
column is weakly-coupled, the right strongly-coupled. 

\subsection{Lie duality}

The only fact present which has not already been described is
the relationship of $\epsilon'$ to $\epsilon$. Recall that
Lie duality operated classically (at zeroth order in $\epsilon$)
in the left-hand column. At first order ({\em i.e.\ }from
semiclassical calculations)\cite{macka94,deliu94b} it also applied with 
$\epsilon'=-\epsilon$ (with real-coupled particles and
imaginary-coupled solitons). However, on examination of the exact
mass ratios we find that it works to all orders only with
\be\label{S1}
{\epsilon' \over 1 + { h^\vee \over h}\epsilon'}
= -{\epsilon \over 1 + {\tilde{h}^\vee\over h}\epsilon}
\hspace{0.2in} \Rightarrow \hspace{0.2in}
\epsilon' = -{\epsilon\over 1+ {h^\vee+\tilde{h}^\vee\over h}\epsilon} \,,
\ee
where $h$ is the Coxeter number (of $g$ and $g^\vee$), $h^\vee$ is the
dual Coxeter number of $g$, and $\tilde{h}^\vee$ that of $g^\vee$.

To see this we require that breather poles in S-matrices for general
algebras work analogously 
to the $b_n$ case, the only case in which the S-matrices have been 
calculated and thoroughly examined. Our conjecture (which we do not
believe to be too strong, since the form of the poles is highly
constrained) is that for any Lie algebra and any species,
\be\label{mass}
m= 2 M \sin \left({\pi|\alpha|^2\over 2 h \lambda}\right)
\;, \hspace{0.2in}{\rm where}\hspace{0.2in} 
\lambda=-{1\over\epsilon}-{h^\vee\over h} \,;
\ee
$m$ denotes particles, $M$ solitons, and $\alpha$ is the relevant
simple root, with $|$long root$|^2=2$.

A simple illustrative example is given by $g_2$, which is Lie
self-dual. Hence there is only one classical mass ratio, $\sqrt{3}$.
Duality exchanges the long and short roots, and hence also the
heavy and light species of solitons compared with particles.
Under the strong$\leftrightarrow$weak affine duality, 
the particles' exact quantum mass ratio is 
$$
{m_h\over m_l}=2\cos{\pi\over H(\epsilon)}
$$ 
($h$ means `heavy', $l$ `light') with
$$
{1\over H(\epsilon)} = {1\over 6} - {1\over 18} {\epsilon \over 1 + 
{2 \over 3}\epsilon}\,,
$$
which runs to the affine dual ($d_4^{(3)}$) ratio in its 
strong-coupling limit (checked\cite{cho93} in perturbation theory
to order $\beta^4$).
Does this expression also give the exact soliton mass ratio? -
that is, does the $g_2^{(1)}\leftrightarrow d_4^{(3)}$ pair contain
only one exact quantum mass ratio to describe both 
solitons and particles?\footnote{Thanks to Ed Corrigan
for persistently asking me this question.} The answer is no.
Whilst this ratio's strong limit gives the correct ratio for
the $d_4^{(3)}$ soliton masses, we can see from
$\epsilon'\neq\epsilon$ in (\ref{S1}) that its weak value does
{\em not} give the correct $g_2$ soliton mass ratio.

Unfortunately, whilst some of the soliton S-matrices 
in the $g_2^{(1)}\leftrightarrow d_4^{(3)}$ pair have been
calculated\cite{takacs97}, they are not enough to prove (\ref{mass})
for $g_2$, which, with $h^\vee=4,h=6$, yields
\beaa
m_l & = &  2 M_h\sin\left({\pi\over
18\lambda}\right)  \\
m_h & = &  2 M_l\sin\left({\pi\over
6\lambda}\right) \,.
\eeaa
The solitons then have a {\em
new} mass ratio whose $\epsilon\rightarrow 0$ limit is $\sqrt{3}$,
$$
{M_h \over M_l} =2\cos{\pi\over H(\epsilon')} \,,
$$
with $\epsilon'$ related to $\epsilon$ by (\ref{S1}). In the strong
limit $\epsilon\rightarrow\infty$ this ratio runs not to the
$d_4^{(3)}$ ratio $2\cos(\pi/12)$ but to $\sqrt{2}$.

Is there some overall duality scheme into which these results
among the masses fit? Suppose we seek a duality between the 
$(\hat{g},\epsilon)$ and $(\hat{g}',\epsilon')$ theories, in which,
for some
functional dependence of $\epsilon'$ on $\epsilon$, solitons and
particles are exchanged. Such a relationship would of course take no 
account of excited solitons or higher breathers, and so might seem
unlikely.
Consideration of the classical masses
requires $\hat{g}=g^{(1)}, \hat{g}'=g^{\vee(1)}$: that is, Lie
duality. For $g_2$ we seek a relationship
\begin{eqnarray*}
\epsilon & \leftrightarrow & \epsilon' \\
{M_h/M_l} & = & {m_h/m_l}   \\
{m_h/m_l} & = & {M_h/M_l}   \\
{M_h/m_l} & = & {m_h/M_l}   \\
{m_h/M_l} & = & {M_h/m_l} 
\end{eqnarray*}
The first two are what we considered above: they require
\be\label{R3}
2\cos{\pi\over H(\epsilon')} = 2\cos{\pi\over H(\epsilon)}.
\ee
For all to be satisfied we require
\begin{eqnarray}
&\sin\left({\pi\over 18\lambda'}\right) \sin\left({\pi\over 6\lambda}\right)  
=  {1\over 4}&
\label{R1} \\[0.1in]
&\sin\left({\pi\over 6\lambda'}\right) \sin\left({\pi\over 18\lambda}\right)  
=  {1\over 4} &
\label{R2}
\end{eqnarray}
(We find that (\ref{R3}) is equivalent to (\ref{R1})/(\ref{R2}).)
But there is no $\epsilon'=f(\epsilon)$ which 
solves (\ref{R1}) and (\ref{R2}),
and hence no exact quantum Lie duality in this form.

This analysis generalizes to the $b_n\leftrightarrow c_n$ case.
The $b_n$ spectrum is extracted from the S-matrices\cite{gande95c},
and gives (\ref{mass}) with $h^\vee=2n-1,h=2n$:
\begin{eqnarray*}
{M_a\over M_n} & = & {\sin{a\pi\over H} \over \sin{n\pi\over H}} 
\hspace{0.3in}(a=1,2,..,n-1)\\[0.1in]
m_a & = & 2 M_a \sin {\pi\over 2n\lambda}  \\[0.1in]
m_n & = & 2 M_n \sin {\pi\over 4n\lambda}  \\[0.1in]
\Rightarrow {m_a\over m_n} & = & 2 \sin{a\pi\over H}\,,
\end{eqnarray*}

For $c_n$, the known particle mass ratios
$$
{m_a\over m_n}  =   {\sin{a\pi\over H'} \over \sin{n\pi\over H'} }
$$
(again for $a=1,2,..,n-1$) will lead via (\ref{mass}), now with 
$h^\vee=n+1,h=2n$, 
to soliton masses satisfying
\begin{eqnarray*}
m_a & = & 2 M_a \sin {\pi\over 4n\lambda'} \\[0.1in]
m_n & = & 2 M_n \sin {\pi\over 2n\lambda'}\, ,
\end{eqnarray*}
giving in turn
$$ 
{M_a\over M_n} = 2 \sin{a\pi\over H'}
$$
(where our priming of quantities simply denotes their
association with $c_n$ rather than $b_n$).

We now seek $\epsilon'=f(\epsilon)$ such that
\begin{eqnarray*}
(b_n,\epsilon) & \leftrightarrow & (c_n,\epsilon') \\
{M_i/M_j} & = & {m_i/m_j}   \\
{m_i/m_j} & = & {M_i/M_j}   \\
{M_i/m_j} & = & {m_i/M_j}   \\
{m_i/M_j} & = & {M_i/m_j}    
\end{eqnarray*}
for all $i,j=1,2,..,n$.
The first of these is what we considered earlier, and is satisfied
if (\ref{S1}) holds, giving a  weak$\leftrightarrow$weak relationship.

The full set is satisfied if 
\begin{eqnarray}
&\sin\left({\pi\over 4n\lambda'}\right) \sin\left({\pi\over 2n\lambda}\right)  
=  {1\over 4} &
\label{RR1} \\[0.1in]
&\sin\left({\pi\over 2n\lambda'}\right) \sin\left({\pi\over
4n\lambda}\right)   
=  {1\over 4} &
\label{RR2}
\end{eqnarray}
holds. 
In the semiclassical limit (eqns (6.5,6.6) of [32]) it was pointed out 
that full $b_n\leftrightarrow c_n$ Lie duality requires a
strong$\leftrightarrow$weak relationship,
$$ 
\beta^2 \beta'^2 = 8 h^2
$$
(that paper contained a misprint), which is the same as the leading
term obtained from either (\ref{RR1}) or (\ref{RR2}),
$$
{\pi^2\over 2 h^2 \lambda \lambda'} = {1\over 4} .
$$
However, at all-orders (\ref{RR1}) and (\ref{RR2}) again 
have no solution for arbitrary $\epsilon$, and we find {\em no} 
Lie duality in this form. There will be solutions for certain
fixed $\epsilon,\epsilon'$, but we have no indication that
these would be significant.

\subsection{Strongly-coupled solitons}

As we pointed out, such a duality would seem too naive given
the presence in the spectrum of many other excited states.
However, as the coupling becomes strong ($\epsilon$ large
and negative) such states disappear from the spectrum.
Strongly-coupled solitons are problematic classically and
semiclassically, but there is no problem with examining
their proposed exact S-matrices at such values of the coupling.
How do solitons behave at strong coupling $\epsilon\rightarrow\infty$?
Their S-matrices are of course non-diagonal, and we can never hope
to identify soliton and particle S-matrices. However, in the infinite
coupling limit the leading term in the S-matrix is just the identity: 
might the solitons behave like particles in this limit? 

If we were to use the discussion
at the beginning of this section as a guide, we might expect to see
a relation between $(g^{\vee(1)}, \epsilon = - 1/\alpha)$ solitons and 
$(g^{(1)\vee}, \alpha - {h^\vee+\tilde{h}^\vee\over h})$ particles
(where $\alpha$ is small and positive).
In the $a_2^{(1)}$ case  (to pick the simplest example), the relation 
would be between $(a_2^{(1)}, - 1/\alpha)$ solitons and 
$(a_2^{(1)}, \alpha-2)$ particles, which does not occur: the particle
$S_{11}$, for example,
is certainly not unity at $\epsilon=-2$, and indeed is so only in 
the weak- and strong-coupling limits. The unlikeliness
of (\ref{S1}) in simply-laced cases (where the rigid mass ratios give
us no guidance on how $\epsilon'$ and $\epsilon$ are to be related) leads
us to suspect that it is only an artefact.
If the solitons are to behave like weakly-coupled particles
at all, their masses, of order $m/\epsilon$, seem to
make it more relevant to investigate mid-ranges of the coupling,
$\epsilon\sim 1$.

\subsection{An observation on soliton S-matrices}

In our initial scheme, the $g^{\vee(1)}$
quantum solitons carry representations of $U_q(g^{\vee(1)\vee})$ 
non-local charges (and thus the $g^{(1)\vee}$ solitons, 
$U_q(g^{(1)})$ charges). Both contain a $U_q(g)$ subalgebra, although
the multiplets form different representations of it\cite{gande95c}.

Specializing to $g=c_n$, we find that we are comparing 
$a_{2n-1}^{(2)}$ with $c_n^{(1)}$ YBE solutions, and
these share a very similar algebraic structure, only differing
in certain algebraic parameters, and hence 
in the form of dependence on the quantum group spectral parameter $x$.
That is,  making the change $(n+1,2n)\mapsto(2n,2n-1)$ in the
$c_n^{(1)}$ solutions yields those associated with $a_{2n-1}^{(2)}$.
Unfortunately this cannot be achieved by a transformation on
$\epsilon$: if we effect the composite Lie-affine transformation 
on $\epsilon$ implicit in our scheme we also rescale the exponents 
in $x$ and $q$. This is inevitable:
the change in $\epsilon$ exchanges strong and weak couplings,
whereas the $(n+1,2n)\mapsto(2n,2n-1)$ observation relates 
S-matrices which are both at weak coupling.

\subsection{Folding non-local charges}

Whereas quantum $b_n^{(1)}$ soliton multiplets
carry representations of
$U_q(a_{2n-1}^{(2)})$ charges, classically the $b_n$ solitons
are `folded' from $d_{n+1}^{(1)}$: that is, they are $d_{n+1}^{(1)}$
solitons invariant under an automorphism of the $d_{n+1}^{(1)}$
Dynkin diagram.  The quantum $d_{n+1}^{(1)}$ soliton multiplets represent 
$U_q(d_{n+1}^{(1)})$ charges, and linear combinations of these,
invariant under a twisting automorphism, form a
$U_q(d_{n+1}^{(2)})$ subalgebra. To describe the quantum $b_n^{(1)}$ 
solitons, however, we would need to add roots and so multiply charges,
and would expect to
find a $U_q(a_{2n-1}^{(2)})$ subalgebra of $U_q(d_{n+1}^{(1)})$. 
It would be useful to understand how the two are related.

\subsection{Further directions}

All we have done here is to set up a few easy targets to
shoot down. Perhaps the overall point is that we should
seek to understand strongly-coupled solitons on their
own terms rather than expect straightforward dualities with particles.
Certainly, more investigation of the known soliton S-matrices
at strong and, especially, at mid-range coupling is needed.
However, we should also stand back
a little and look for alternative ways forward. 

Of primary importance
is the generalization of the fusing rule to the full spectrum.
In simply-laced cases, and for solitons only or particles only,
mass ratios do not renormalize and Dorey's rule is
therefore rigid. Any extension of it, either to nonsimply-laced 
cases or to a rule for the rich arrangement\cite{gandethesis} 
of couplings {\em between} solitons and particles or breathers,
requires a flexible, $\beta$-dependent and therefore intrinsically
quantum construction. Steps have been made towards both
former\cite{oota97} and latter\cite{iska94}, 
but the full construction remains some way off.

At a grander level, the goal is an algebraic scheme in
which $g^{(1)}$, $g^{\vee(1)}$, $g^{(1)\vee}$ and 
$g^{\vee(1)\vee}$ all make appearances. Such a scheme
would have eventually to incorporate all of our results,
of course, but let us point out some of those likely to
be more salient.
First, the Coxeter number plays a central role in Dorey's construction,
and so the incorporation of the flexible Coxeter number $H$
is a correspondingly crucial step.
Second, an emerging fact\cite{evans97}  is that when (the non-commuting,
non-additive, non-integer spin) quantum group charges exist, so do 
Wilson's local charges 
(which are commuting, additive and have spins equal to the exponents).
Unfortunately, the constructions are always rather model-specific;
what we would like is a  model-independent construction incorporating
both.
Finally, in the background remains our lack of understanding of quantum
group representations: why Dorey's rule applies
to their tensor products remains a central unexplained fact.

Remarkably, such a construction may be emerging in the theory of
deformed W-algebras\cite{frenkel97}. 
Much remains conjectural, but it is becoming clear
that different limits of these deformations yield the centres of
different quantized
affine algebras, dual in various ways, and that the W-algebras can
be used to construct representations of these. Discovering the facts
of ATFTs within this picture is likely to make an interesting quest.

\pagebreak
{\bf Acknowledgments}

I should like to thank G\'erard Watts for helpful comments,
and Pembroke College, Cambridge, for a Stokes Fellowship.

\end{document}